# Spiking dynamics of frequency up-converted field generated in continuous-wave excited rubidium vapours

Alexander M. Akulshin[1,2*], Nafia Rahaman[1], Sergey A. Suslov[3], Dmitry Budker[2,4], and Russell J. McLean[1]

[1] *Optical Sciences Centre, Swinburne University of Technology, PO Box 218, Melbourne 3122, Australia*
[2] *Helmholtz Institute, Johannes Gutenberg University, D-55128 Mainz, Germany*
[3] *Department of Mathematics, Faculty of Science, Engineering and Technology, Swinburne University of Technology, Melbourne, Australia*
[4] *Department of Physics, University of California, Berkeley, CA 94720-7300, USA*
*Corresponding author: aakoulchine@swin.edu.au*

We report on spiking dynamics of frequency up-converted emission at 420 nm generated on the $6P_{3/2}$-$5S_{1/2}$ transition in Rb vapour two-photon excited to the $5D_{5/2}$ level with laser light at 780 and 776 nm. The spike duration is shorter than the natural lifetime of any excited level involved in the interaction with both continuous and pulsed pump radiation. The spikes at 420 nm are attributed to temporal properties of the directional emission at 5.23 μm generated on the population-inverted $5D_{5/2}$-$6P_{3/2}$ transition. A link between the spiking regime and cooperative effects is discussed. We suggest that the observed stochastic behaviour is due to the quantum-mechanical nature of the cooperative effects rather than random fluctuation of the applied laser fields. © 2018 Optical Society of America

http://dx.doi.org/10.1364/AO.99.099999

## 1. INTRODUCTION

Nonlinear parametric and nonparametric processes in atomic media generate new optical fields with substantially different wavelengths to those used for excitation [1]. The frequency conversion of continuous wave (cw) laser light with a power of several tenths of a milliwatt into directional blue, ultraviolet and mid-infrared (IR) radiation in alkali vapours continues to be an active topic of research [2-14].

In Rb atoms, for example, combined stepwise and two-photon excitation to the $5D_{5/2}$ energy level, as shown in Figure 1a, produces population inversion on the dipole-allowed $5D_{5/2}$-$6P_{3/2}$ transition because the spontaneous decay probability from the $6P_{3/2}$ level is higher than that from the $5D_{5/2}$ to $6P_{3/2}$ levels [15]. If both the excitation rate and the atom number density in the pencil-shaped interaction region are sufficiently high, mirrorless lasing occurs at 5.23 μm [16] due to the process of amplified stimulated emission (ASE) [17, 18]. Mixing of the generated forward-directed mid-IR radiation with the copropagating applied laser fields produces collimated blue light (CBL) at 420 nm by the process of parametric four-wave mixing (FWM) in the direction that satisfies the phase-matching condition [4, 19].

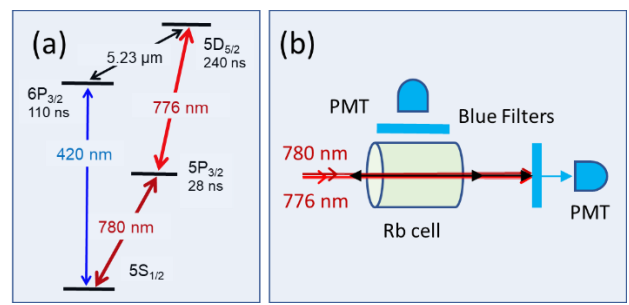

**Fig. 1.** (a) Energy levels of Rb atom involved in two-photon excitation with resonant laser light at 780 and 776 nm to produce new fields at 5.23 μm and 420 nm. (b) Optical scheme of the experiment.

Although this approach to frequency conversion has attracted a lot of attention, the dynamics of the process have not been considered in detail. When analysing the intensity of frequency up-converted light at 420 nm with a sub-μs temporal resolution, we find that this directional emission is extraordinarily noisy. It appears that the relative intensity noise of the CBL is much larger than the noise of both the applied laser fields.

Here we present the results of our experimental investigation of the temporal behaviour of CBL generated by FWM in two-photon excited

Rb vapours. We attribute this dynamic to temporal properties of the mid-IR radiation at 5.23 μm that is involved in the FWM process despite the fact that its temporal characteristics are not directly analysed since in our laboratory fast detectors are available only for the visible and near-IR spectral regions.

We find that the CBL temporal behaviour resembles the spiking in solid-state lasers [20, 21]. Also, we discuss a link between the temporal dynamics of the new field and cooperative effects in population-inverted media [22, 23], an extensive study of which over past decades was initiated by R. H. Dicke in his seminal paper [24].

## 2. EXPERIMENTAL SETUP

Our experimental study of the temporal behaviour of the frequency up-converted field is conducted using a standard experimental arrangement for new optical field generation described in recent papers [9, 16]. The optical scheme of our experiment is shown in Figure 1b. Frequency tuneable radiation at 780 nm and 776 nm is delivered by a Toptica TA-100 tapered amplifier and a homemade extended cavity diode laser, respectively. The upper limit for the short-term laser linewidth of both the lasers is 2 MHz. This has been estimated using optical heterodyning with an auxiliary laser. The linearly polarized radiation of the lasers is combined on a non-polarizing beam splitter (BS) with an approximately 50:50 split ratio. Radiation from each laser is weakly focused before being combined on the BS. Polarization of the combined bi-chromatic beam is controlled with a polarizer and a quarter-wave plate.

The new-field generation occurs in a 5-cm long sealed cell having a glass body and sapphire windows. The cell is surrounded by a μ-metal foil to reduce the ambient magnetic field by about 3-4 times. The full width half maximum (FWHM) beam diameter inside the cell at 780 and 776 nm is approximately 0.4 mm. The maximum power of the pump laser radiation at 780 and 776 nm before entering the cell is 14 and 5 mW corresponding to intensities of 11 and 4 W/cm$^2$, respectively.

The cell contains a natural mixture of Rb isotopes with no buffer gas. The cell is heated to 65 - 90 °C, so that the Rb atom number density $n$ varies from $4 \times 10^{11}$ to $2.0 \times 10^{12}$ cm$^{-3}$. The design of the heater allows the cell windows to be kept approximately 5 °C hotter than the rest of the cell, avoiding undesirable metal condensation on the windows.

The cell with nearly parallel windows is tilted by approximately 10 degrees to avoid overlap of the reflected and incident radiation, as this can affect the characteristics of the generated fields [10].

The fixed optical frequency of the 780 nm laser is controlled using the auxiliary laser scanning across the Rb-D$_2$ absorption line. A precise reference for evaluating the fixed frequency of the 780 nm laser is provided by a beat signal produced by mixing radiation from the TA-100 laser and the auxiliary 780 nm laser in comparison with the Doppler-free polarization spectroscopy spectrum realized in an additional room-temperature Rb cell.

Sub-μs temporal resolution of the CBL and side-detected blue fluorescence is achieved using available photomultipliers (PMTs), Hamamatsu R136 and R446, which are characterized by a rise time of 2 ns and a transit time of 22 ns, and Tektronix oscilloscopes with detection bandwidths of 100 and 500 MHz.

## 3. EXPERIMENTAL RESULTS

### A. Spiking dynamics of directional emission at 420 nm

Figure 2a shows a typical single-scan and 16-scan averaged spectral profiles of CBL recorded with sub-μs temporal resolution. The single-scan trace shows that the directional emission at 420 nm consists of partially overlapping temporal spikes. Detection of CBL with a larger number of averages or with a lower detection bandwidth leads to complete blurring of the spikes suggesting they have a stochastic origin.

As was shown previously, CBL possesses a pronounced threshold-type dependence on the atom number density $n$ and the applied laser power [4, 6, 7, 10]. At our typical experimental conditions, such as the laser power at 776 nm $P_{776}$ =2 mW and $n \approx 1.2 \times 10^{12}$ cm$^{-3}$, the CBL threshold occurs approximately at $P_{780} = P_{TH} = 0.6$ mW.

We first analyse the CBL intensity fluctuations with fixed frequencies for both applied cw laser fields at 780 and 776 nm. Figure 2b shows an example of variations of CBL intensity, recorded over a 10 ms interval. The relative standard deviation (RSD) of the directional blue emission (RSD = $\sigma/\mu \times 100\%$, where $\sigma$ and $\mu$ are the sample standard deviation and the mean value of the sample data set, respectively) varies from scan to scan between 10 and 60%, while for the applied laser radiation at both the wavelengths it is much smaller, RSD = (0.9 ± 0.2) %.

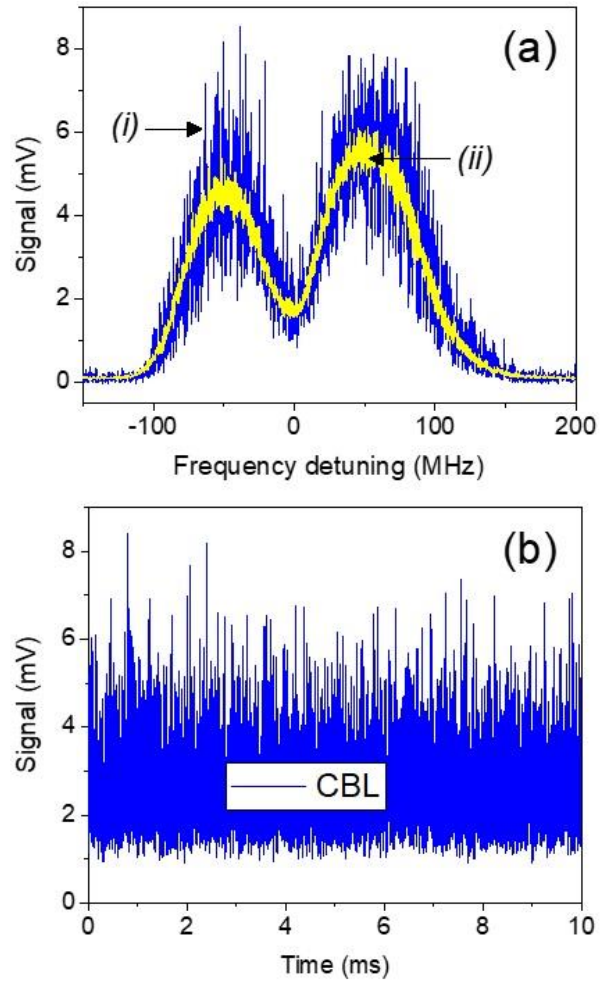

**Fig. 2.** (a) Single-scan (i) and 16-scan averaged (ii) intensity profiles of the CBL as a function of the 776 nm laser frequency that is 50 Hz scanned in the vicinity of the $^{85}$Rb 5P$_{3/2}$(F'=4)-5D$_{5/2}$ transition. The fixed frequency 780 nm laser is tuned to the $^{85}$Rb 5S$_{1/2}$(F=3)-5P$_{3/2}$(F'=4) transition. The atom number density in the cell is $n \approx 1.0 \times 10^{12}$ cm$^{-3}$; the applied laser powers at 780 and 776 nm are 1.5 and 2.0 mW, respectively. (b) CBL intensity variations, when the fixed frequencies of the 780 nm and 776 nm lasers are tuned to the 5S$_{1/2}$(F=3)-5P$_{3/2}$(F'=4) and 5P$_{3/2}$(F'=4)-5D$_{5/2}$ transitions, respectively, to maximise CBL intensity.

We find that despite the cw laser pumping, the directional blue emission consists of distinct irregular pulses on a zero background if the excitation rate is not high. A typical sequence of CBL pulses detected at approximately 30% above the CBL threshold condition is shown in Figure 3. Here and in what follows the fixed frequency lasers at 780 and 776 nm are tuned to the $5S_{1/2}(F=3)-5P_{3/2}(F'=4)$ and $5P_{3/2}(F'=4)-5D_{5/2}$ transitions to maximise the CBL intensity. Zooming in on a few consecutive peaks, we find that the duration of a spike can be shorter than 15 ns (FWHM).

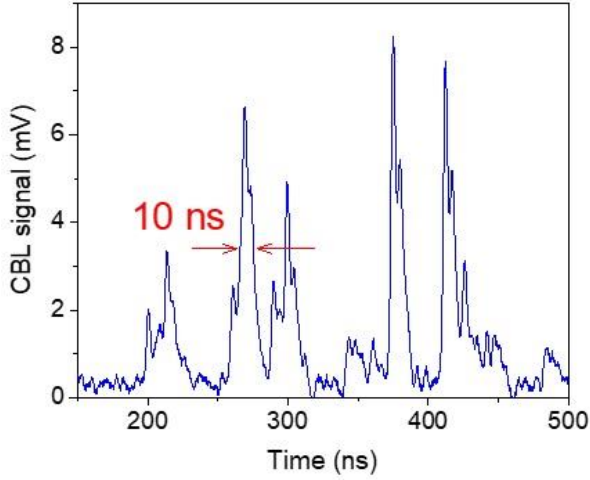

**Fig. 3.** Typical sequence of CBL spikes detected using an oscilloscope with a 500 MHz detection bandwidth. The fixed frequency lasers are tuned to the $5S_{1/2}(F=3)-5P_{3/2}(F'=4)$ and $5P_{3/2}(F'=4)-5D_{5/2}$ transitions to maximise the CBL intensity. The applied cw pump power $P_{776}$ =2.0 mW and $P_{780}$=0.8 mW, while $n \approx 5.5 \times 10^{11}$ cm$^{-3}$.

As the excitation rate is increased, for example by increasing the laser pump power, the amplitude of the CBL pulses becomes larger and time intervals between pulses decrease. At a certain excitation level, pulses begin to overlap. This causes the minimum CBL level to exceed the dark noise level of the detection system. However, large spikes of CBL, which are well above average, still occur as shown in Figure 2b.

### B. Stochastic character of CBL spikes

To demonstrate that the CBL spikes do not occur due to random fluctuations of the applied laser light, we compare temporal dependences of two CBL fields generated simultaneously under similar experimental conditions.

First, we check the level of amplitude correlation of the signals from the two available PMTs when they detect the same CBL divided by a non-polarizing beam splitter. The calculated Pearson coefficient $r$ [25], which is a measure of the linear correlation between two variables $x$ and $y$, that shows how they co-vary on average with time and defined as

$$r = \frac{\sum_j (x_j - \langle x \rangle)(y_j - \langle y \rangle)}{\sqrt{\sum_j (x_j - \langle x \rangle)^2 \sum_j (y_j - \langle y \rangle)^2}},$$

where $\langle x \rangle$ and $\langle y \rangle$ are the means, shows a considerable correlation ($r = 0.513 \pm 0.018$). The correlation is limited by random dark noise contributions of the PMTs. The dark noise signals (no light applied) show a very weak correlation ($r = 0.011 \pm 0.004$).

Next, we compare temporal dependences of two CBL beams produced in spatially separated regions inside the cell and simultaneously detected, as shown in Figure 4a. The separation between the regions, approximately 3 mm, is larger than their diameters. The total laser power of bichromatic pumping in the separation regions, as well as the power ratio between the components at 780 and 776 nm, differ by less than 10%. In addition, to ensure that the overlap between the components is as similar as possible, the distance between the cell and the BS is made the same for both the bichromatic beams.

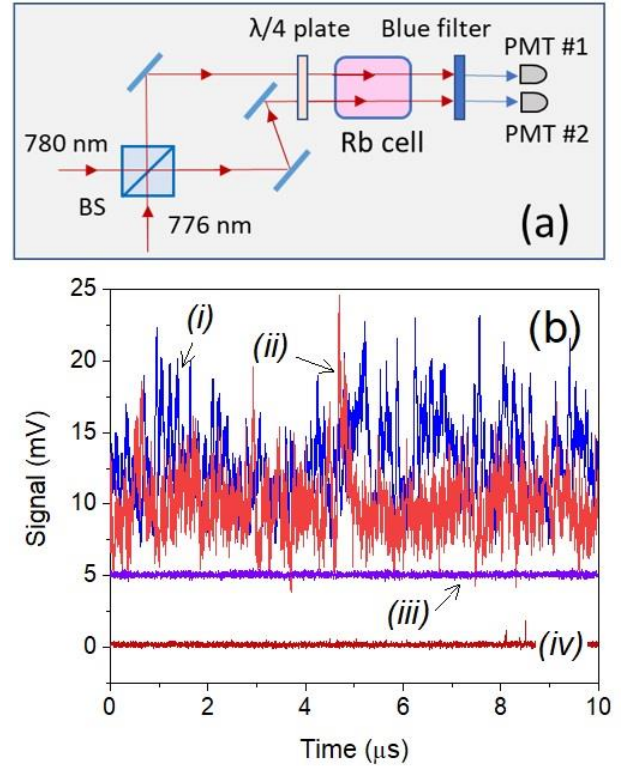

**Fig. 4.** a) Optical setup for comparing two CBL fields generated in spatially separated regions. The parallel bichromatic beams inside the cell are 3 mm apart, while their diameters are approximately 0.4 mm. BS: non-polarizing beam splitter. b) Curves (i) and (ii) show typical single-scan CBL temporal profiles, while curves (iii) and (iv) represent dark-noise signals recorded by PMT#1 and PMT#2, respectively. Curve (iii) is shifted vertically by 5 mV for clarity. The applied cw pump powers at 780 and 776 nm are 4.8 mW and 2.0 mW, respectively. The atom number density in the cell is $n \approx 1.3 \times 10^{12}$ cm$^{-3}$.

Figure 4b demonstrates typical single-scan temporal profiles of CBL simultaneously recorded by both PMTs #1 and #2, as well as their dark-noise signals, during a randomly selected time interval. Despite the average levels of the signals being close, (13.1±2.8) mV and (9.9±2.3) mV, the temporal profiles differ significantly; in particular, maxima or minima of the CBL fields detected by PMTs #1 and #2 do not coincide in time. However, qualitatively, they have the same character and very close RF spectra.

The Pearson coefficient $r$ calculated for curves (i) and (ii) in Figure 4b has a small negative value ($r = -0.059$), however, it varies from shot to shot. The averaged $r$ for a set of curves recorded at the same experimental conditions shows a weak anticorrelation ($r = -0.050 \pm 0.026$). These values are significantly different from the coefficient $r$ calculated for signals when the PMTs detect the same CBL field as mentioned above. These observations suggest that CBL spikes are not triggered by random fluctuations in the pump radiation; otherwise, the intensities of the two CBL fields generated in the spatially-separated regions should be correlated in time.

## C. Pulse excitation

For a more systematic study of the temporal characteristics of CBL, we use pulsed laser excitation since this provides a reliable reference point in time. For this, the output beam from the Toptica TA-100 at 780 nm is frequency shifted and 100% intensity modulated using an acousto-optic modulator (AOM) before combining on the BS with the cw laser light at 776 nm. The AOM is controlled with rectangular pulses with a repetition frequency of up to 2 MHz. The produced optical pulses are characterized by an approximately 170 ns rise/fall time determined by the AOM bandwidth. The FWHM of the pulses can be varied from 0.16 to 25 µs.

Now we again consider single rather than dual-CBL field generation. When Rb vapour is excited with pulsed light at 780 nm and cw light at 776 nm, CBL is also generated as a sequence of single or partly overlapping pulses on a zero background.

The generated CBL appears as a single spike if the laser pulse duration is shorter than 200 ns, as shown in Figure 5. Figure 5a demonstrates three single-scan CBL profiles each containing four consecutive pulsed excitations. The blue light pulse amplitude varies from shot to shot by up to 80% of the average level. The FWHM of the CBL pulse duration is approximately 43 ± 6 ns, which is significantly shorter than the applied laser pulse ($\simeq$170 ns). The CBL temporal profile averaged over 16 scans fits well to a Gaussian profile, as Figure 5b demonstrates. Its FWHM is approximately 55 ns, which is larger than the width of single-scan spikes. This occurs because the spike appearance time relative to the applied pulse fluctuates, broadening the averaged profile.

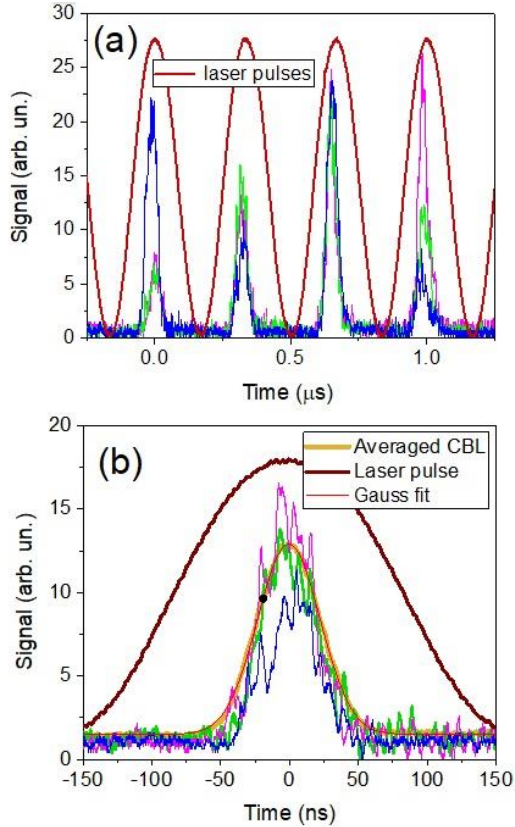

**Fig. 5.** (a) Three single-scan CBL temporal profiles generated with four excitation pulses at 780 nm (FWHM $\simeq$170 ns) and cw laser light at 776 nm. The detection bandwidth is 100 MHz. (b) Three single-scan CBL temporal profiles and 16-shot averaged profile with a 38.7 ns-long Gaussian fit. The averaged excitation laser pulse at 780 nm is shown for reference.

With longer laser pulses, the temporal dependence of the CBL signal becomes more complex, as shown in Figure 6. After an initial short and strong spike, the CBL intensity drops and then revives apparently randomly. The initial spikes are also characterised by a higher reproducibility in their occurrence time, duration and shape compared with the subsequent spikes, which are usually smaller.

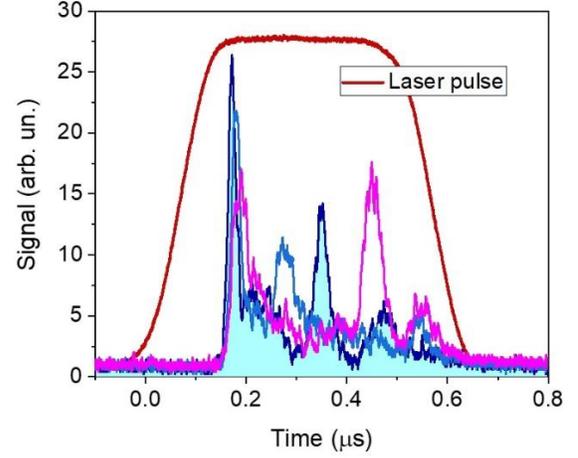

**Fig. 6.** Three consecutive single-scan temporal profiles of CBL generated with a longer excitation pulse at 780 nm (FWHM is approximately 0.5 µs) and cw laser light at 776 nm. The Rb atom number density in the cell is $n \approx 1.3 \times 10^{12}$ cm$^{-3}$. The detection bandwidth is 100 MHz.

Despite the apparent randomness in the appearance of subsequent spikes in the case of long pump pulses, CBL signal averaging reveals a quasiperiodic structure in its temporal behaviour, as demonstrated in Figure 7. The averaged curve shows that the time intervals between intensity peaks are close. We find that this period depends on several experimental parameters, such as the atom number density $n$, the applied laser intensity and its polarization, as well as on laser frequency detuning from the corresponding one-photon transitions. At our experimental conditions the period of oscillations varies from 50 to 200 ns. At optimum experimental parameters the oscillations can last up to 3 µs after the start of pump pulses of corresponding duration.

Such temporal behaviour of CBL resembles relaxation oscillations, which are common in lasers with a high optical gain and a relatively low-finesse optical cavity [20, 21]. The results of our study of this oscillating appearance of CBL spikes will be reported elsewhere.

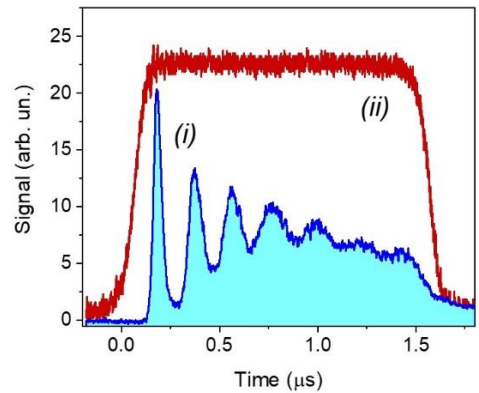

**Fig. 7.** Curve (i) shows eight-time averaged temporal profile of CBL generated with (ii) 1.5 µs-long laser pulses at 780 nm (peak power $P_{780}$ = 1.8 mW) and cw radiation at 776 nm ($P_{776}$ = 2.8 mW), while $n \approx$ 1.3×10$^{12}$ cm$^{-3}$. Detection bandwidth is 100 MHz.

A typical evolution of the averaged CBL temporal profiles with increasing applied laser intensity is illustrated in Figure 8. Just above the CBL power threshold, for example, if $P_{780} \leq 1.1 P_{TH}$, CBL pulses show large variations of their appearance time, although it is more likely that the pulse occurs near the centre of 1.1 μs-long laser pulse. After averaging, the directional blue emission appears as a weak and relatively broad pulse (FWHM ≃100 ns). At higher applied power that results in a higher excitation rate to the $5D_{5/2}$ level, the first CBL spike becomes stronger and it appears earlier in time. Also, subsequent spikes become clearly observable at $P_{780} > 1.5\ P_{TH}$, however, the first spike is noticeably shorter (FWHM ≃ 30 ns) than the subsequent ones.

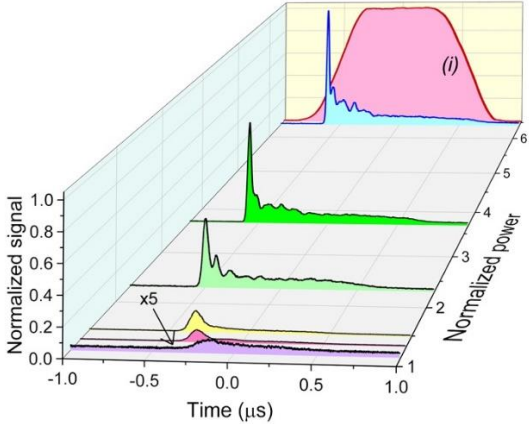

**Fig. 8.** Averaged temporal profiles of CBL generated at different power of the pulsed laser light at 780 nm and constant cw power at 776 nm, $P_{776}$ = 2.9 mW, while $n \approx 1.3 \times 10^{11}$ cm$^{-3}$. The pump pulses are normalized on its threshold value $P_{TH}$ = 0.6 mW. Curve (i) shows temporal profile of the 1.1 μs-long pump pulse. The origin of the time scale is set at the centre of the pulse. The fixed frequency 780 nm and 776 nm lasers are tuned to maximise the CBL intensity.

**D. CBL and blue fluorescence**

We note that the temporal characteristics of CBL and side-detected blue fluorescence are quite different, as shown in Figure 9.

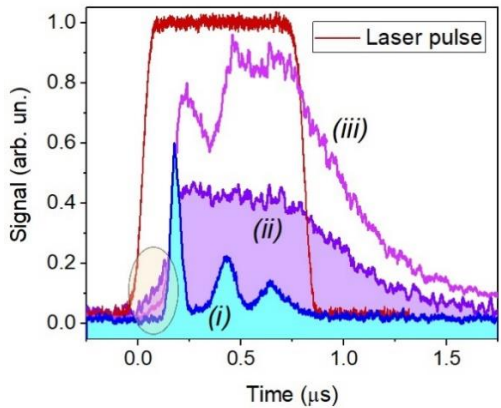

**Fig. 9.** 16-shot averaged temporal profiles CBL and side-detected blue fluorescence in Rb vapours ($n \approx 1.3 \times 10^{12}$ cm$^{-3}$) excited by approximately 1 μs-long laser pulse at 780 nm ($P_{780}$ = 3 mW) and cw radiation at 776 nm ($P_{776}$ = 2.0 mW). Curve (i) shows the CBL profile, while curves (ii) and (iii) represent temporal variations of blue fluorescence detected near the entrance window and in the centre of the cell, respectively.

Fluorescence at 420 nm can be detected at least 100 ns earlier than the CBL, as shown in a highlighted area; however, its rise is less steep. Also, we find that the temporal behavior of blue fluorescence depends on the position of the PMT along the Rb cell. Fluorescence detected in the proximity of the entrance window rises to a maximum level then slowly declines during the pump pulse, while fluorescence detected from the central part of the cell shows pronounced intensity variations, which are to some extent anti-synchronized with the CBL spikes. This means that the fraction of atoms in the $6P_{3/2}$ levels affected by the process responsible for the CBL field generation is growing with the pump propagation inside the cell. After the pump pulse ends, both fluorescence signals decay steadily and curves (ii) and (iii) can be approximated by an exponential function with the time constant τ = 380 ± 30 ns.

## 4. DISCUSSION

It is unlikely that the process of nonlinear wave mixing of cw optical fields is itself responsible for the observed spiking behaviour of CBL. We attribute it to temporal features of the directional mid-IR emission at 5.23 μm. We believe that the temporal behavior of this radiation is determined by *cooperative effects* [22-24] occurring on the population-inverted $5D_{5/2}$-$6P_{3/2}$ transition. If so, the observed spiking emission at 420 nm is the result of transferring the temporal properties of the mid-IR field to the frequency up-converted field through the FWM process.

It is well-known that cooperative effects (CE) result in dramatic modifications of radiative dynamics of a system of excited emitters. Under certain conditions, the inverted system can relax to the lower level radiating much faster than an ensemble of independent emitters. The duration of the CE pulse is shorter than the natural lifetime of the upper level of the inverted transition and is inversely proportional to the number of inverted atoms. Moreover, since synchronization of individual atoms emitting radiation with initially random phases takes a certain time, called the delay or induction time, the emitted CE pulse is delayed relative to the pump pulse.

In the near-IR optical domain CE were first reported on cascading transitions in two-photon and one-photon excited sodium [26] and rubidium vapours [27]. Later, short pulses of visible radiation at 420 nm on the 6P-5S transition were obtained with laser-cooled [28] and hot Rb atoms [29, 30] excited to the 5D levels. The observed time-delayed pulses of sub-natural duration were attributed to the CEs, since their properties match their most distinctive characteristics. However, as was mentioned in [28], the radiation at 420 nm was observed under conditions when CE emission was not expected. Indeed, the population inversion that is required to establish cooperative emission is unlikely to have been prepared on the 6P-5S transition. The observations were explained by cooperative cascade emission [31], which suggests the presence of CE emission at 5.23 μm. However, to the best of our knowledge this mid-IR field has not been directly analysed with a high temporal resolution.

In our experiment, we find that the typical duration of individual spikes observed for both cw and pulsed excitation lies between 15 and 50 ns, which is shorter than the natural lifetime of both the $5D_{5/2}$ and $6P_{3/2}$ levels. In addition, the approximately 100 ns-delayed start of the CBL spike with respect to the side-detected fluorescence is an indication of the delayed emission at 5.23 μm. These features, along with directionality of the CBL and mid-IR emission that was previously reported [4, 9, 11] and confirmed here, match the characteristic properties of CE radiation.

It is interesting to note that some temporal features of the blue emission generated in our experiment are somewhat similar to the conical emission that can be generated in liquids, glasses and atomic media due to parametric wave-mixing. In particular, pulses of the conical emission might be also significantly shorter than the pump laser pulse as was demonstrated in Sr vapour excited to the 5 $^1P_1$ level with near-resonant laser pulses at 480 nm [32].

## 5. MODELLING

The stochastic nature of the emitted spikes reflects their quantum-mechanical origin, arising due to fluctuations of the vacuum field in the process of spontaneous decay [33]. Proper theoretical treatment of any stochastic behaviour generally requires a sophisticated formalism, which is beyond the scope of this paper. However, we find that a simple model based on rate equations for the photon numbers and level populations for a three-level atom is able to reproduce some features of the temporal behaviour of the frequency up-converted field observed with signal averaging.

In the model excitation occurs from the ground state $|a\rangle$ to the excited level $|b\rangle$ with excitation rate $R$ (Fig. 10 a). The excited atoms spontaneously decay to the ground state $|a\rangle$ and the intermediate level $|c\rangle$ at rate $\gamma_{ba}$ and $\gamma_{bc}$, respectively. Population inversion on the $|c\rangle$ - $|b\rangle$ transition, established if $\gamma_c > \gamma_{bc}$, is responsible for optical amplification and under some conditions for mirrorless lasing. Figure 10 shows a numerical solution for the stimulated radiation on the $|b\rangle$-$|c\rangle$ transition. The temporal behaviour of the emitted directional light is consistent with some observed features. For example, the lasing intensity $I$ grows, the intervals between the peaks are reduced, while the width of the peaks become smaller with increasing pump rate $R$. In addition, the appearance of directional light is delayed relative to the excitation pulse, and this delay decreases with higher excitation. We note that these observations are also in qualitative agreement with the results of numerical simulation performed using a more complex model based on a semiclassical approach [29]. The modelling shows that the actual shape of the excitation pulse has a strong influence on the emission of the blue light.

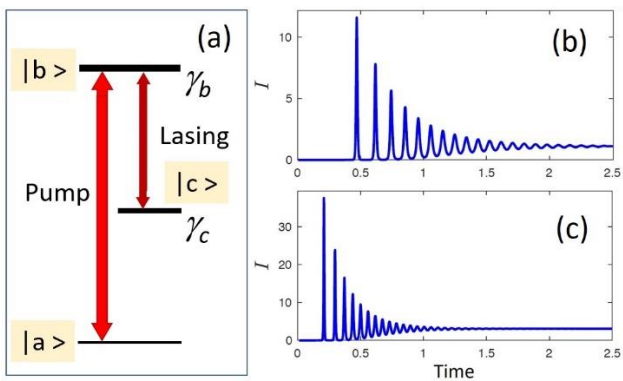

**Fig. 10.** (a) Three-level scheme used for modelling. Figures (b) and (c) demonstrate temporal behaviour of the lasing intensity $I$ in arbitrary units modelled for the case of step-wise excitation $R$=0.2 and $R$=0.4, respectively, which starts at t = 0. The time scale is normalized on $1/\gamma_b$.

## 6. CONCLUSION

In this work, we have focused our attention on the temporal dynamics of directional emission at 420 nm obtained in Rb vapours two-photon excited to the $5D_{5/2}$ level with resonant radiation at 780 and 776 nm.

We show that despite cw and quasi-cw two-photon excitation of Rb atoms, the directional blue emission generated by parametric FWM exhibits a pronounced spiking behaviour. The spike duration is shorter than the natural lifetime of any excited level involved in the interaction. We attribute this to the temporal characteristics of the directional emission at 5.23 μm generated on the population-inverted $5D_{5/2}$-$6P_{3/2}$ transition, although we are not able to directly analyse this emission with high temporal resolution. This spiking regime on the inverted transition obtained with cw excitation could be investigated directly using fast mid-IR detectors. Alternatively, appropriate excitation schemes in Rb or other alkali atoms could be used.

A link between the spiking regime and the cooperative effects is discussed. The observed stochastic appearance of the CBL spikes is explained by the quantum-mechanical nature of the cooperative effects.

**Funding.** This work has been supported in part by the US Office of Naval Research Global under grant N62909-16-1-2113.

**Acknowledgment**. We acknowledge fruitful discussions with W. Gawlik, I. Novikova, A. Wickenbrock, and M. Chekhova. We are also grateful to A. Sidorov for providing the Toptica TA-100 laser.